\documentclass[prd,12pt]{revtex4}
\usepackage{graphics}
\usepackage{epsfig}

\newcommand\ba{\begin{eqnarray}}
\newcommand\ea{\end{eqnarray}}

\newcommand{\br}[1]{\left( #1 \right)}

\newcommand{\GeV}{~\mbox{GeV}}

\begin{document}

\title{Two-photon mechanism of production of scalar mesons at colliders}

\author{M.~K.~Volkov}
\email{volkov@theor.jinr.ru}
\affiliation{%
Joint Institute for Nuclear Research, Dubna, Russia
}%
\author{E.~A.~Kuraev}
\email{kuraev@theor.jinr.ru}
\affiliation{%
Joint Institute for Nuclear Research, Dubna, Russia
}%

\author{Yu.~M.~Bystritskiy}
\email{bystr@theor.jinr.ru}
\affiliation{%
Joint Institute for Nuclear Research, Dubna, Russia
}%

\begin{abstract}
    The cross sections of the scalar meson $f_0(980)$, $a_0(980)$ and $\sigma(600)$
    production in collision of electron and positron beams were calculated.
    The two-photon decays of the scalar mesons, obtained in the framework of the
    Nambu-Jona-Lasinio model, were used. The quark and meson loops
    were taken into account.
\end{abstract}

\maketitle


The nature of the scalar mesons $f_0(980)$, $a_0(980)$ and $\sigma(600)$
is the subject of intensive investigation during the last years
(see \cite{Volkov:2009mz} and references therein).
In theoretical description of these mesons the different models are used.
In one of these models scalar mesons were considered as quark-antiquark states
\cite{Volkov:2009mz,Volkov:2008ye}.
In other models \cite{Achasov:1987ts,Achasov:2008ut}
these scalar mesons were represented as four-quark states.
There are models where scalar mesons are considered as a mixture
of quark-antiquark and four-quark states \cite{Gerasimov:2003tn}.
Also the model exists where scalar mesons were described as
kaon molecules \cite{Weinstein:1990gu,Branz:2008ha}.

Here we will use the model where scalar mesons are considered as
quark-antiquark states in the framework of Nambu-Jona-Lasinio (NJL) model.
In this model, we take into account both the quark and the meson loops
\cite{Volkov:2009mz,Volkov:2008ye}.

Let us note that with the scalar mesons (such as $\sigma$ and $f_0$)
which have quantum numbers similar to vacuum quantum numbers
a set of anomalous phenomena could be connected.
For example, there is some problem with the explanation of the
production ratio of $K$ and $\pi$ mesons in the
proton-antiproton annihilation in the $~^1S_0$ and $~^1P_0$ states,
namely in the $~^1S_0$ state of the initial proton-antiproton pair the
production ratio of $K$ and $\pi$ mesons is comparable with each other.
However, in the $~^1P_0$ state -- this ratio differs by a few orders of magnitude.

The aim of this paper is to point out the possibility to obtain large statistics of
scalar meson production at electron-positron colliders.
With the use of the \cite{Volkov:2009mz}
two-photon decay widths of scalar mesons
the Brodsky-Kinoshita-Terazawa formula \cite{Brodsky:1970vk} gives:
\ba
    \sigma_S
    &=&
    \frac{8\alpha^2 \Gamma_S}{M_S^3}
    \ln^2\br{\frac{s}{m_e^2}}
    f\br{\frac{M_S^2}{s}}, \\
    f\br{z} &=&
    \br{2+z}^2 \ln\br{\frac{1}{z}} -
    2\br{1-z}\br{3+z}, \label{BKT}
\ea
where $\Gamma_S$, $M_S$ are the two-photon decay width and mass
of scalar meson (eg. $f_0$, $a_0$ or $\sigma$),
$s$ is the invariant mass of the initial electron-positron pair.

Let us note that formula (\ref{BKT}) allows us to estimate
these cross sections using experimental data for two-photon
decay widths of the scalar mesons \cite{Amsler:2008zzb}.

\begin{figure}
\includegraphics[width=0.8\textwidth]{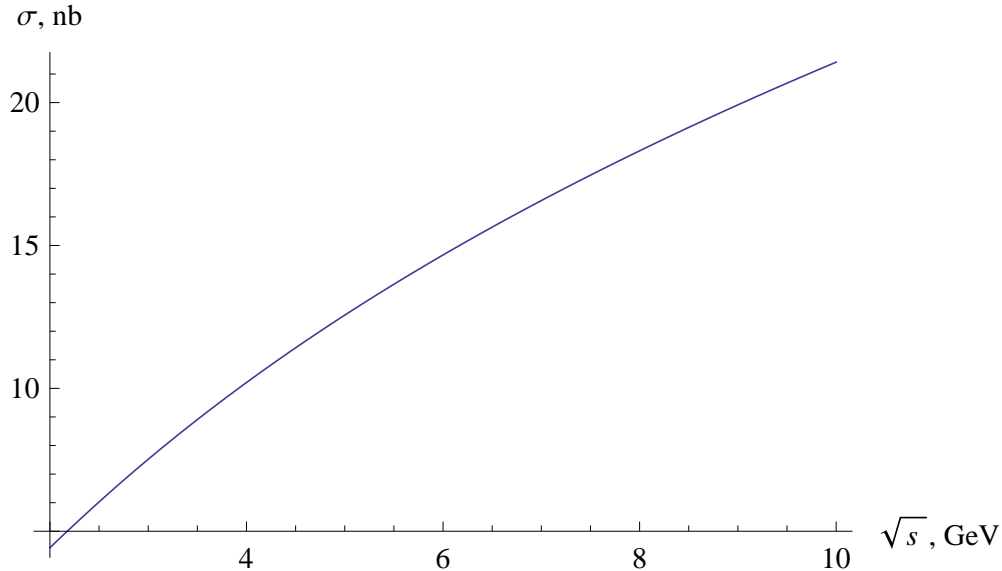}
\caption{The total cross section of $\sigma(600)$ meson production at
electron-positron collision as a function of total invariant mass $s$.
\label{Fig1}}
\end{figure}

\begin{figure}
\includegraphics[width=0.8\textwidth]{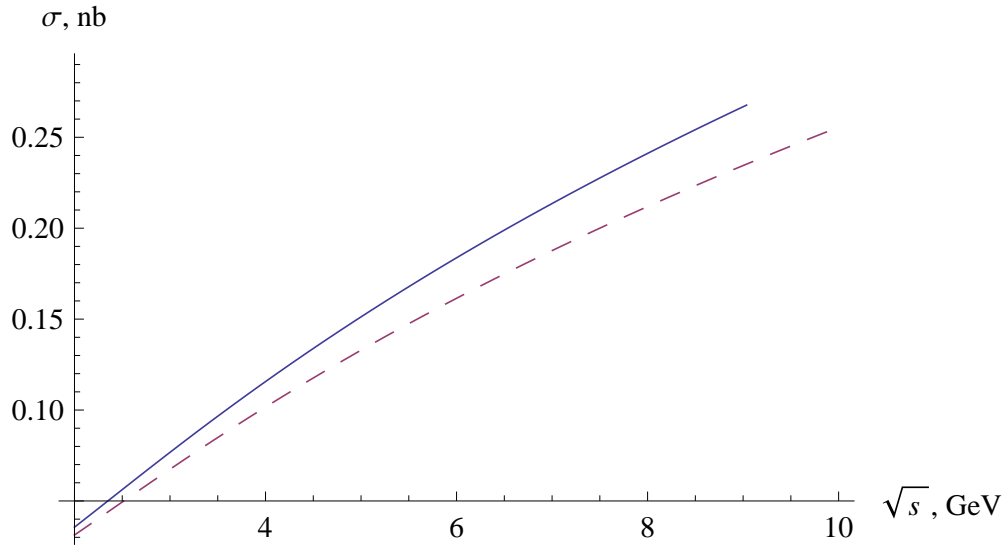}
\caption{The total cross sections of $f_0(980)$ (solid curve) and $a_0(980)$
(dashed curve) mesons production at
electron-positron collision as a function of total invariant mass $s$.
\label{Fig2}}
\end{figure}

The cross sections of the
$f_0$, $a_0$ and $\sigma$ meson production are given on Fig.~\ref{Fig1}, \ref{Fig2}.
It is seen that in the energy range $3 < \sqrt{s} < 10 \GeV$ the cross sections
of meson production are of several nanobarns.
Thus, the experimental facilities with luminosities of an order of
$L \sim 10^{-33-34}~\mbox{cm}^{-2} \mbox{s}^{-1}$
(BES (Beijin), CLOE (Frascati), VEPP-2 (Novosibirsk))
can be considered as scalar meson factories
with production of $10^3 - 10^4$ mesons per day.

Let us note that this mechanism is the dominant one for
scalar meson production at electron-positron colliders.
The production of mesons through annihilation mechanism
is suppressed by a factor of $\br{m_e^2 M_S^2/s^2}$.

Thus, on the basis of these facilities it is possible to test different models
for explanation of the nature of scalar mesons.


\end{document}